\documentclass[epj]{svjour}
\usepackage{graphics}
\usepackage{pslatex}
\begin{document}
\title{Intercalation and High Temperature Superconductivity of
Fullerides}
\author{A.~Bill\inst{1}
 \and V.Z.~Kresin\inst{2}% etc
% \thanks is optional - remove next line if not needed
%\thanks{}%
}                     % Do not remove
%
%\offprints{}          % Insert a name or remove this line
%
\institute{Paul Scherrer Institute, Condensed Matter Theory, 5232
Villigen PSI, Switzerland \and Lawrence Berkeley Laboratory,
University of California at Berkeley, CA 94720, USA}
\date{Received: 23 January 2002}
% The correct dates will be entered by Springer
%
\abstract{
 Intercalation of polyatomic molecules into a superconductor can
drastically affect the properties of the compound. A mechanism
responsible for a large increase in $T_c$ for such systems is proposed.
It explains the recent remarkable observation of high $T_c$
superconductivity in the hole-doped C$_{60}$/CHX$_3$ (X$\equiv$Cl,Br)
compounds and the large shift in their $T_c$ upon Cl$\to$Br
substitution. The increase in $T_c$ is due to contribution to the
pairing arising from the interaction of electrons with the vibrational
manifold of the molecule. The proposed mechanism opens up the
possibility to observe a site-selective isotope effect. We also
suggest that intercalating CHI$_3$ would further increase the critical
temperature to $T_c\simeq 140$K.
 % end of PACS codes
} %end of abstract
\maketitle

\noindent
This paper is concerned with the impact of intercalation by polyatomic
molecules on the properties of superconductors. The study has been
motivated by the recent remarkable observation of high $T_c$ caused by
intercalation of CHBr$_3$ (bromoform) and CHCl$_3$ (chloroform) molecules
into hole-doped fullerides \cite{schoen1} (see also description of
\cite{schoen1} in Ref.~\cite{service}). The discovery of high
temperature superconductivity \cite{schoen1} in these systems ($T_c =
117$K for C$_{60}$/CHBr$_3$; $T_c = 80$K for C$_{60}$/CHCl$_3$) has
attracted a lot of attention and raises the fundamental question about
the nature of this phenomenon.\\
The usual superconducting fullerides (see ,e.g., the
reviews \cite{hebard,gunnarsson}) are chemically electron-doped
compounds (e.g., Rb$_3$C$_{60}$, K$_3$C$_{60}$). It is believed that their
superconductivity is due to the coupling of electrons to the
intramolecular vibrational modes. The vibrational spectrum spreads
over a broad region ($\Omega_L\simeq 270$cm$^{-1} \to \Omega_H\simeq
1500$cm$^{-1}$). Strictly speaking, all modes contribute to the pairing
but the question of which region of the vibrational spectrum plays a
major role is still controversial (see, e.g. \cite{gunnarsson}).\\
A new exciting development was described in \cite{schoen2}; with the
use of gate-induced doping, a hole-doped fulleride compound was
created without chemical dopant. As a result a major increase in $T_c$
to 52K(!) was observed. This is caused by the different properties of
the valence band relative to the conduction band such as the density
of states. Recently, this technique was complemented by
intercalating the molecules CHBr$_3$ and CHCl$_3$ into the crystal
\cite{schoen1}. These molecules are placed in regions near the
C$_{60}$ units. This has resulted in a dramatically higher $T_c$
again: $T_c \simeq 117$K(!). It was indicated in \cite{schoen1} that
the change in lattice spacing may be responsible for the increase in
$T_c$.  This is an interesting scenario which needs to be studied in
detail. Here we propose another additional mechanism. It is of
interest in its own right; furthermore, it can provide for a
quantitative explanation of the drastic increase in $T_c$ observed in
\cite{schoen1}.\\
It is very essential to note that the experimentalists \cite{schoen1}
observed
not only a remarkable high $T_c$ for the C$_{60}$/CHX$_3$ compound
(X$\equiv$ Br,Cl), but also a large difference in the value of $T_c$
for X$ \equiv$ Br {\it vs.} X$\equiv$ Cl. Whereas $T_c\equiv
T_C^{Br}\simeq 117$K was observed for X$\equiv$Br, the value of $T_c$
for X$\equiv$ Cl turned out to be much smaller ($\simeq 80$K),
although still very high. The structures of the bromoform and
chloroform molecules are similar and therefore a microscopic
explanation of the strong change in the critical temperature is also
of definite interest.

 The picture we propose is that the intercalated molecules themselves
actively, and importantly, participate in the pairing. The internal
vibrational modes of the molecules provide for an additional
attraction which leads to higher $T_c$. Such a mechanism was
considered by one of the authors in \cite{kresin}. Therefore, our major
focus is on the vibrational spectra of these molecules.

Let us start with the parent hole-doped fulleride ($T_c$ = 52K). The order
parameter for this superconductor is described by the equations:
\begin{eqnarray}
\label{Eq1}
\Delta(\omega_n) Z(\omega_n) &=& \\
&& \hspace*{-.75cm}\pi T \sum_{\omega_{n'}}
\left[\lambda_1 D(\omega_n-\omega_{n'}; \tilde{\Omega}_1) -
\mu^\star\right] \frac{\Delta(\omega_{n'})}{|\omega_{n'}|}\nonumber\\
\label{Eq2}
Z(\omega_n) &=& 1 + \lambda_1 \pi T \sum_{\omega_{n'}}
D(\omega_n-\omega_{n'}; \tilde{\Omega}_1)
\frac{\omega_{n'}}{|\omega_{n'}|}
\end{eqnarray}
Here $\lambda_1$ is the electron-phonon coupling constant,
$\tilde{\Omega}_1$ is the characteric (average) intramolecular
vibration frequency\\ ($\tilde{\Omega}_1 = <\Omega^2>_{\rm
  intra}^{1/2}$; see, e.g., Ref.~\cite{gunnarsson}),
$D = \tilde{\Omega}_1^2\left[\tilde{\Omega}_1^2 +
  (\omega_n-\omega_{n'})^2\right]^{-1}$ is the phonon Green's
function, $\omega_n = (2n+1)\pi T$, $T=T_c$, $Z$ is the renormalization
function, and $\mu^\star$ is the Coulomb pseudopotential. We are
employing the temperature Green's functions formalism (see,
e.g. \cite{AGD}).

Eqs.~(\ref{Eq1}), (\ref{Eq2}) correlate the value of $T_c$ with the
values of the parameters $\tilde{\Omega}_1$, $\lambda_1$,
$\mu^\star$. It is natural to assume that hole-doped
fullerides \cite{schoen2} differ from electron-doped compounds obtained
by chemical doping in the magnitude of the coupling constant
$\lambda_1$, whereas the values of the characteristic phonon frequency
$\tilde{\Omega}_1$ and $\mu^\star$ should remain similar.
Following Ref.~\cite{gunnarsson}, let us take as an estimate
$\tilde{\Omega}_1\simeq 0.1$eV  and $\mu^\star = 0.15$. Then,
the value of $T_c$ = 52K observed in \cite{schoen2} implies the 
value $\lambda_1 \approx 0.75$. We would
like to emphasize that if we were to adopt a different value of
$\tilde{\Omega}_1$ (e.g., $\tilde{\Omega}_1 =80$meV, or even 50meV),
this would affect some intermediate parameters (see below), but the
final result, namely the value of $T_c$ for the C$_{60}$/CHBr$_3$
compound, would not be affected in any noticeable way.

In the intercalated C$_{60}$ compound, polyatomic molecules CHX$_3$
(X$=$Cl,Br) are added at interstitial sites. (see, e.g., Fig.~1 in
Ref.~\cite{service}). As indicated above, their presence is found to
result in two phenomena: 1.~A remarkable high $T_c$ ($T_c\simeq 117$K)
for X$\equiv$Br; 2.~A very high, but significantly lower $T_c$
($\simeq 80$K) for intercalation by chloroform (X$\equiv$Cl).

The molecules have a relatively complex vibrational spectra (see,
e.g., Ref.~\cite{herzberg}), but it can be modeled as an
additional phonon mode. Then, instead of equations
(\ref{Eq1}, \ref{Eq2}) we have:
\begin{eqnarray}
\label{Eq3}
\Delta(\omega_n) Z(\omega_n) = \hspace*{6.cm}\\
\pi T \sum_{\omega_{n'}}
\left[\sum_{i=1,2}\lambda_i D(\omega_n-\omega_{n'}; \tilde{\Omega}_i) -
\mu^\star\right] \frac{\Delta(\omega_{n'})}{|\omega_{n'}|}\hspace*{.45cm}
\nonumber\\
\label{Eq4}
Z(\omega_n) = 1 +  \pi T \sum_{i=1,2}\lambda_i \sum_{\omega_{n'}}
D(\omega_n-\omega_{n'}; \tilde{\Omega}_i)
\frac{\omega_{n'}}{|\omega_{n'}|}\hspace*{.5cm}
\end{eqnarray}
The values of $\lambda_1$ and $\tilde{\Omega}_1$ correspond to the
parent, non-inter\-ca\-la\-ted state [see Eq.~(\ref{Eq1})]. We assume that
the value of $\mu^\star$ is also essentially unchanged, although the
presence of molecules and their electrons might lead to an effective
decrease in $\mu^\star$ (see below).

 The presence of the second term ($\sim \lambda_2$) in the sum on the
 right-hand side of Eq.~(\ref{Eq3}) leads to an increase in $T_c$
 relative to that of the parent compound ($T_c^o\simeq 52$K).
The magnitude of this increase depends on the values of
$\tilde{\Omega}_2$ and $\lambda_2$.

As is known, the value of the coupling constant $\lambda$ is described
by the McMillan relation \cite{McMillan1}:
\begin{eqnarray}
\label{Eq5}
\lambda = \frac{\nu <I^2>}{M <\Omega^2>},
\end{eqnarray}
where $\nu$ is the density of states, $I$ is the matrix element, and
$M$ and $\Omega$ describe the vibrational spectrum. We will show that
the differences in the vibrational spectra of the CHX$_3$ molecules
can account quantitatively for the observed large difference in $T_c$
upon chemical substitution (Cl$\to$Br). Let us now turn to
quantitative analysis.

The vibrational spectrum of the CHCl$_3$ molecule (see,
e.g. \cite{herzberg}) contains five major modes which can contribute
to the pairing. These modes are $\Omega^{(1)Cl} \simeq
  262$cm$^{-1}$, $\Omega^{(2)Cl} \simeq 366$cm$^{-1}$, $\Omega^{(3)Cl}
  \simeq 668$cm$^{-1}$ , $\Omega^{(4)Cl} \simeq 761$cm$^{-1}$,
  $\Omega^{(5)Cl} \simeq 1216$cm$^{-1}$ (the notation $\Omega^{Cl}$
  means that we consider the chlo\-ro\-form mo\-le\-cu\-le; the modes
  $\Omega^{(1)Cl}$, $\Omega^{(4)Cl}$, $\Omega^{(5)Cl}$  are
  degenerate).
All these modes are Raman active and can couple to the conduction
electrons \cite{gunnarsson,herzberg}.

 Consider now the CHBr$_3$ molecule. It has the same configuration and a
similar set of modes, but all of them are noticeably softened.
Indeed, according to Ref.~\cite{herzberg},
  $\Omega^{(1)Br} \simeq 154$cm$^{-1}$,\\$\Omega^{(2)Br} \simeq
  222$cm$^{-1}$, $\Omega^{(3)Br} \simeq 539$cm$^{-1}$ ,
  $\Omega^{(4)Br} \simeq 656$cm$^{-1}$, $\Omega^{(5)Br} \simeq
  1142$cm$^{-1}$.
As a result, it  can be expected that the well-known "softening" mechanism
(cf., e.g. \cite{matthias}) manifests itself.  That is, $T_c$ is
expected to increase as a result of the   substitution.

 The effective coupling constants [see Eq. (5)] also contain the mass
($M$) factor. It is crucial that all these modes in
  CHX$_3$ molecule correspond, mainly, to the motion of the light
  carbon ion relative to the heavy X$_3$ (X$\equiv$ Cl, Br) unit.  As a 
result, $M\approx M_C$  and the difference in coupling strength
between the two compounds is mainly determined by the aforementioned
frequency shift ($\lambda_i \sim \Omega_i^{-2}$). Note that the
  hydrogen stretching mode is not essential since it lies much higher
  in energy and, if desired, can be singled out by isotope
  substitution.

Let us consider the C$_{60}$/CHCl$_3$ system first. Using the above
mentioned vibrational modes taken from literature \cite{herzberg}, we
determine the average 
frequency to be $\tilde{\Omega}_2\equiv \Omega_2^{Cl}=85$meV (care
should be taken of the degeneracy of the modes). Performing a
numerical calculation with Eqs.~(\ref{Eq3}),(\ref{Eq4}) and the derived
frequency $\tilde{\Omega}_2$, we
obtain that the observed value $T_c = 80$K \cite{schoen1} implies the
value $\lambda_2^{Cl}\approx 0.2$. Such a value is small and perfectly
realistic, confirming that the observed rise upon intercalation can be
assigned to the action of the added intramolecular modes.

 Let us now substitute Cl$\to$Br and compare C$_{60}$/CHCl$_3$ and
C$_{60}$/CHBr$_3$. This comparison is an important step in our
analysis. As was stressed above, the total coupling constant
$\lambda_2^{Br}$ changes relative to $\lambda_2^{Cl}$ primarily as a
result of the vibrational frequency shifts. Using the literature
values for the vibrational frequencies $\Omega_i^{Br}$ and
$\Omega_i^{Cl}$ (see above), we calculate that $\lambda_2^{Br}\approx
2.65 \lambda_2^{Cl}$. Therefore, the "softening", indeed, leads to a
noticeable increase in the coupling constant.

Since $\lambda_2^{Cl}\approx 0.2$ (see above), we obtain
$\lambda_2^{Br}\approx 0.55$. Using this value for $\lambda_2^{Br}$
and the frequency $\Omega_2^{Br}\simeq 70$meV infered from the above
vibrational spectra, we solve Eqs.~(\ref{Eq3}) and (\ref{Eq4}) and
obtain $T_c \simeq 110K$ for the C$_{60}$/CHBr$_3$ compound. It is
important to emphasize once more that during this last step the value
of $T_c^{Br}$ has been obtained without any additional adjustable
parameter. Table \ref{tab:table1} summarizes our numerical results.
\begin{table}
\caption{Calculation of $\lambda_2^{Cl}$, $\lambda_2^{Br}$ and $T_c$
  for C$_{60}$/CHBr$_3$. $\lambda_1$: coupling to intramolecular
  vibrations of C$_{60}$, $\lambda_2$: additional coupling to the
  vibrational modes of CHCl$_3$ or CHBr$_3$. The coupling constant
  $\lambda_2$ for CHBr$_3$ is fixed by the other set of parameters
  (see text).}
\label{tab:table1}       % Give a unique label
% For LaTeX tables use
\begin{tabular}{llll}
\hline\noalign{\smallskip}
&hole-doped C$_{60}$ & C$_{60}$/CHCl$_3$ & C$_{60}$/CHBr$_3$\\
\noalign{\smallskip}\hline\noalign{\smallskip}
$\lambda_1$ & 0.75 & 0.75 & 0.75\\
$\lambda_2$ & 0 & 0.2 & 0.55\\
$T_c$ [K]& 52$^a$ & 80 & 110\\
\noalign{\smallskip}\hline
$^a$ Ref.~\cite{schoen2}
\end{tabular}
% Or use
%\vspace*{5cm}  % with the correct table height
\end{table}
The derived value $T_c\simeq 110$K  is close to the experimentally
observed one $T_c\simeq 117$K \cite{schoen1}.
Since our calculation made use of some approximations, the agreement
can be considered as rather good.

Refs.~\cite{schoen1} and \cite{schoen3} also describe
the study of electron-doped systems. We first note that doped
electrons are filled into the conduction band made of orbitals with
$T_{1u}$ symmetry (at the $\Gamma$ point), whereas doped holes fill in
the valence band which has $H_{1u}$ symmetry (see, e.g.,
Ref.~\cite{chancey}.) A group theoretical analysis shows that in the
former case, electrons couple to intramolecular vibrations of the
C$_{60}$ cluster belonging to the five-fold degenerate $h_g$
irreducible representation. In the hole-doped case, on the other hand, the
higher symmetry of the valence band allows for coupling of holes to
additional vi\-bra\-tio\-nal modes. Both $h_g$ and the four-fold $g_g$
vibrations are Jahn-Teller active in hole-doped systems. One can
therefore expect that the overall coupling is larger for holes than for
electrons, thereby enhancing $T_c$.

The procedure described above for hole-doped C$_{60}$/CHX$_3$ can be
applied to the electron-doped case and reproduce the results obtained
in Ref.~\cite{schoen1} as well. Indeed, from
Eqs. (\ref{Eq1},\ref{Eq2}) we conclude that the observed value $T_c =
11$K \cite{schoen3} is reached for $\lambda_1 \approx 0.5$. Since the
frequencies of the $h_g$ and $g_g$ modes are in the same range
\cite{onida}, we take the same value of $\tilde{\Omega}_1$ for
electron- as for hole-doped systems. From Eqs.~(\ref{Eq3},\ref{Eq4})
we further obtain the experimentally observed $T_c = 18$K for
C$_{60}$/CHCl$_3$ with an additional coupling $\lambda_2^{Cl} \approx
0.07$. The scaling relation $\lambda_2^{Br}\approx 2.65
\lambda_2^{Cl}$ then leads to $\lambda_2^{Br}\approx 0.2$. As a result
we obtain $T_c \approx 30$K for electron-doped C$_{60}$/CHBr$_3$. This
is in good agreement with the experimental value $T_c\simeq 26$K
\cite{schoen1}.

We have focused on the vibrational spectra of the intercalated
molecules. Note that an additional contribution to the pairing could also
come from the Coulomb interaction and the corresponding polarization
of the electronic systems of the mo\-le\-cu\-les, virtual transitions
between electronic levels (see Ref. \cite{geilikman}). This additional
attraction would correspond to the electronic energy scale and can be
described as an effective decrease of $\mu^\star$
(cf. Ref. \cite{bill}). Note that the polarizability
$\alpha^{Br}\equiv \alpha_{CHBr_3}$ is much larger than $\alpha^{Cl}$
\cite{karna}. This question will be considered elsewhere.
Another factor indicated in \cite{schoen1} is the increase in "ionic"
radius upon intercalation and Cl$\to$Br substitution. However, recent
experiments indicate that lattice stretching resulting from the
intercalation cannot account for the increase in $T_c$ \cite{huq}. It
thus remains an open question whether this effect by itself is capable
of producing such a giant impact on $T_c$.

We propose also the following experiment directly related to our
approach. Namely, one can observe site selective isotope effect by
substituting $^{12}$C$\leftrightarrow^{13}$C on the intercalated
molecules. Such substitution will affect $T_c$ because the vibrational
manifold of these molecules contribute to the pairing. From
Eqs.~(\ref{Eq3},\ref{Eq4}) and the experimentally observed shift in
$\Omega^{(j)Cl}$ $j=1,\dots,5$ for CHCl$_3$ \cite{clark} we obtain
that $\Delta T_c = T_c(^{12}\rm{C}) - T_c(^{13}\rm{C}) \simeq
-0.2$K. Though small, such a shift can be measured experimentally
(cf.~\cite{franck}). The corresponding isotope coefficient\\
$\alpha_C^{Cl}\equiv -(M_C/\Delta M_C)(\Delta T_c/T_c)$ is of order
$0.03$. If, in addition, one also substitutes
$^{35}$Cl$\leftrightarrow^{37}$Cl a shift in $T_c$ of about $\sim
-0.35$K can be observed.

A more precise numerical treatment would require proper use of
tunneling spectroscopy, similar to that described in
Ref. \cite{McMillan2}, in order to obtain the accurate form of
$\alpha^2(\Omega) F(\Omega)$ \footnote{Note that the analysis of
  $\alpha^2F$ has to be made with considerable care, as the
  vibrational manifold of the intercalated molecules partially
  overlaps with that of the C$_{60}$ intramolecular modes.}.
Our aim was to identify the mechanism responsible for the strong
increase in $T_c$ for the intercalated compounds. We argued that this
mechanism involves the vibrational manifolds of the intercalated
molecules. We demonstrated that this mechanism can naturally lead to
high values of $T_c$, as observed in Ref.\cite{schoen1}.

Finally, we would like to propose to carry out the same experiment as in
Ref.~\cite{schoen1} with the compound C$_{60}$/CHI$_3$. The
corresponding modes are $\Omega^{(1)I} \simeq
  110$cm$^{-1}$, $\Omega^{(2)I} \simeq 154$cm$^{-1}$, $\Omega^{(3)I}
  \simeq 425$cm$^{-1}$ , $\Omega^{(4)I} \simeq 578$cm$^{-1}$,
  $\Omega^{(5)I} \simeq 1068$cm$^{-1}$ \cite{dawson}. Our analysis
shows that such intercalation leads to $T_c \simeq 140$K(!)
Note also that the site selective isotope effect
$^{12}$C$\leftrightarrow^{13}$C for CHI$_3$ molecules has the additional
advantage that iodine has no natural isotopes.

The mechanism of pairing proposed here and connected with the
interaction of electrons with vibrational manifolds of intercalated
molecules can provide for a drastic increase in $T_c$.
The search for new systems which would similarly
permit the intercalation by appropriate polyatomic molecules is an
exciting and promising direction in the quest of new high temperature
superconductors.\\

\noindent{\small We thank B.~Batlogg for sending the manuscript
  (Ref.~\cite{schoen1}) prior to publication.}\\[.3cm]

\noindent{\bf Note added in proof:}\\ 
Recent tunneling measurements \cite{schoen4} have shown the appearance
of additional peaks upon intercalation. This additional structure
corresponds to the vibrational manifold of the molecules, and
Cl$\rightarrow$Br substitution leads to the ``softening''. The data of
Ref.\cite{schoen4} support our theory.

\end{document}